\documentclass{hep99}
\usepackage{epsf}
\begin{document}

\title{\boldmath
 Measurement of the $CP$-violation parameter $\sin(2\beta)$ 
 in $B^0 \rightarrow J/\psi K^0_s$ decays 
 }

\author{G. Bauer%$^1$
\\ \small
(representing the CDF Collaboration)
}

\address{
              Laboratory for Nuclear Science, 
              Massachusetts Institute of Technology, \\
              Cambridge, MA 02139 USA
\\[3pt]
E-mail: {\tt bauerg@fnal.gov}}

\abstract{A sample of $\sim \!\! 400$ 
          $B^0_d/\overline{B}{^0_d} \!\rightarrow\! J/\psi K^0_s$ decays
collected in $\bar{p}p$ collisions by the CDF detector
is used to directly measure the $CP$-violation parameter $\sin(2\beta)$.
We find $\sin(2\beta) = 0.79 ^{+0.41}_{-0.44}$, 
favoring the standard model expectation 
of a large $CP$ violation in this $B^0$ decay mode.
}

\maketitle

%%%%%%%%%%%%%%%%%%%%%%%%%%%%%%%%%%%%%%%%%%%%%%%%%%%%%%%%%%%%%%%%%%%%%
\section{Introduction\label{sec:intro}}

The origin of $CP$ violation has been an outstanding issue
since its discovery 
in $K^0_L \rightarrow \pi^+\pi^-$ decays 35 years ago \cite{KL_CP}.
In 1972, before charm was discovered,
Kobayashi and Maskawa \cite{CKM} proposed that 
quark mixing with 3 (or more) generations was the cause.
In this case, the CKM matrix relating 
the mass and weak eigenstates of quarks
possesses, in general, a complex physical phase
that violates $CP$.
Unfortunately, the $K^0$ has been the only place to study $CP$ violation.
Despite precision $K^0$-studies,
a complete picture of $CP$ violation
is still lacking.

$CP$ tests have encompassed $B$ mesons,
but the violations in
inclusive studies \cite{B_CP}  
are too small (\mbox{$\sim\!\!10^{-3}$}) to as yet detect.
In the '80's it was realized
that the {\it interference} due to mixing of $B^0_d$
decays to the same $CP$ state could show large violations \cite{CarterSanda}.
%that the  mixing  {\it interference} of $B^0_d$
%decays to the same $CP$ state could show large violations \cite{CarterSanda}.

$B^0_d/\overline{B}{^0_d} \!\rightarrow\! J/\psi K^0_s$ is
the ``golden'' mode for large effects, with
little theoretical uncertainty
relating it to the CKM matrix.
A $B^0_d$ may decay directly to $J/\psi K^0_s$,
or it may oscillate into a $\overline{B}{^0_d}$
and then decay to $J/\psi K^0_s$. 
The two paths have
a phase difference, and
the interference results in an asymmetry:
\begin{equation}
{\cal A}_{CP}(t) \equiv \frac{\overline{B}{^0_d}(t)-B^0_d(t) }
                        {\overline{B}{^0_d}(t)+B^0_d(t) } = 
                       \sin(2\beta) \sin (\Delta m_d t),
\label{eq:cp_asym}
\end{equation}
where $B^0_d(t)$ [$\overline{B}{^0_d}(t)$] is the number of
$J/\psi K^0_s$ decays 
at proper time $t$ from mesons 
produced as $B^0_d$
[$\overline{B}{^0_d}$]. % ($t=0$).
${\cal A}_{CP}$ %oscillates 
varies as $\sin (\Delta m_d t)$ because
it is shifted by a $\frac{1}{4}$-cycle 
relative to the $\cos (\Delta m_d t)$ mixing oscillation
by the mixed/unmixed decay
interference.
The am\-plitude is $\sin(2\beta)$, with %where,
$\beta =$ $ \arg(-V_{cd}^{\,}V_{cb}^*/V_{td}^{\,}V_{tb}^*)$
for CKM elements $V_{qq'}$.
$\beta$ is also an angle from the so-called ``unitarity triangle''
of the CKM matrix.

%%%%%%%%%%%%%%%%%%%%%%%%%%%%%%%%%%%%%%%%%%%%%%%%%%%%%%%%%%%%%%%%%%%%%
\section{\boldmath
The $B^0/\overline{B}{^0} \rightarrow J/\psi K^0_s$ sample
\label{sec:sample}}
\begin{figure}[t]
{\centering
  \epsfxsize=16pc % will enlarge or reduce the postscript figures 
\epsfbox{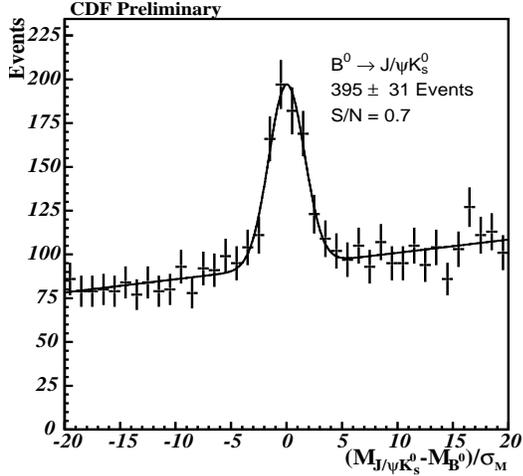}
  \caption{
         Normalized mass distribution $M_N$ (see text) 
         of the $J/\psi K^0_s$ candidates. A Gaussian signal
         plus linear background fit to the data is shown by the curve.
}
\label{fig:mass} }
\end{figure}

\vspace*{-12.0cm}
{ \flushright \large
$\,$\\ %CDF/PUB/BOTTOM/PUBLIC/5071\\
%V2.6
FERMILAB-Conf-99/228-E \\
}
\vspace*{11.05cm}
We exploit the large $B$ cross section at the
Tevatron and obtain a sample of $J/\psi K^0_s$ decays
to measure $\sin(2\beta)$. We start
from the Run I  $J/\psi \!\rightarrow\! \mu^+\mu^-$ sample 
($p_T(\mu)$ above $\sim\!\!1.5\,$GeV/$c$)
of $\sim\!\!\frac{1}{2}$ million events. % (110 pb$^{-1}$).
The $K^0_s \rightarrow \pi^+\pi^-$ reconstruction 
tries all oppositely charged
track combinations (assumed to be pions).
The $p_T(K^0_s)$ must be above 0.7 GeV/$c$,
its decay vertex displaced 
from the $J/\psi$'s by $>\!5\sigma$,
and $p_T(J/\psi K^0_s) > 4.5$ GeV/$c$.
After imposing the $J/\psi$ and $K^0_s$ masses, 
  the fitted $J/\psi K^0_s$ 
  mass $M_{FIT}$ and error $\sigma_{FIT}$ %($\sim 9\;{\rm MeV}/c^2$)
  are used to construct $M_N \equiv (M_{FIT} - M_0)/\sigma_{FIT}$, where
  $M_0$ is the  world average $B^0_d$ mass.
The $M_N$ distribution is shown in Fig.~\ref{fig:mass}.
A likelihood fit
yields $395 \pm 31$ $B^0_d/\overline{B}{^0_d}$'s.

The $\bar{p}p$ collisions spread beyond CDF's
Si-$\mu$vertex tracking detector (SVX), so only about half 
($202 \!\pm\! 18$ vs$.$ $193 \!\pm\! 26$)
of the $J/\psi$'s  have both muons
in the SVX.
The precision life\-time information from the SVX
allows one to make a time-dependent fit 
to Eq.\ref{eq:cp_asym}. However, 
the $CP$ asymmetry remains even when integrated in time;
so although life\-time in\-for\-mation is basically lost in ``non-SVX'' data,
they are still useful. The statistical power
for meas\-ur\-ing $\sin(2\beta)$ is only reduced by $\sim\!{1}/{3}$
for this sub\-sample.
The ``SVX'' subsample was the basis for our previous $\sin(2\beta)$
measurement~\cite{CDFcp}.

%%%%%%%%%%%%%%%%%%%%%%%%%%%%%%%%%%%%%%%%%%%%%%%%%%%%%%%%%%%%%%%%%%%%%
\section{Flavor tagging\label{sec:tagging}}

Observing the asymmetry ${\cal A}_{CP}(t)$
is predicated upon determining
the $b$ ``flavor''---whether the $B$ meson is composed 
of a $b$ or a 
$\bar{b}$ quark---at the time 
of production.
If the initial flavor is correctly tagged
with probability $P$,
then the observed asymmetry 
\endcolumn
\noindent
is attenuated by the ``dilution''
${\cal D} = 2P -1$, {\it i.e.} 
${\cal A}_{CP}^{Obs} = {\cal D} \sin(2\beta)\sin(\Delta m_q t)$.

A method
with tagging efficiency $\epsilon$ yields an error
on $\sin(2\beta)$ which scales as $1/\sqrt{\epsilon {\cal D}^2 N}$ for 
$N$ background-free mesons. Thus,
$\epsilon {\cal D}^2$ measures the effective tagging power.
An analysis can be im\-prov\-ed by using
several taggers, the combined effect is approximately the sum
of the respective $\epsilon {\cal D}^2$'s.

The tagging needs for this analysis are similar to those
employed in $B^0$-$\overline{B}{^0}$ oscillation 
measurements of $\Delta m$. CDF has performed six 
$\Delta m_d$ analyses that demonstrate three types of tagging methods.
The CDF average
$\Delta m_d$ is  $0.495 \!\pm\! 0.026 \!\pm\! 0.025 \,{\rm ps}^{-1}$ 
\cite{CDFMixSum},
which is of similar precision to other experiments and
agrees well with a world average \cite{Artuso}.

We call the first method ``same-side tagging'' (SST),
as it relies on the charge of a particle ``near'' the $B^0$ \cite{Gronau}.
The idea is simple.
A $\bar{b}$ quark forming  a $B^0_d$
combines with
a $d$  in the hadronization, %fragmentation, 
leaving a $\bar{d}$.
To make a charged pion, the $\bar{d}$
combines with
a $u$ making a $\pi^+$. 
Conversely, a $\overline{B}{^0_d}$ will be as\-so\-ci\-ated with a $\pi^-$.
Correlated pions also arise from 
$B^{**+} \!\rightarrow B^{(*)0} \pi^+$
decays.\footnote{A CDF 
$B^{**}$ analysis of $\ell D^{(*)}$ data found
the fraction of $B_{u,d}$ mesons arising as
$B^{**}$ states to be $0.28 \pm 0.06 \pm 0.03$~\protect\cite{Dejan}.
} Both sources have the same correlation, 
and are not distinguished 
here.

The SST tag is the candidate track with the smallest momentum
transverse to the $B$+Track momentum.
A valid track candidate 
must be
within $\Delta R \!=\! \sqrt{(\Delta\eta)^2\!+\!(\Delta\phi)^2} \!\le\! 0.7$
of the $B$,
have $p_T \!>\! 400$ MeV/$c$,
reconstructed in the SVX, 
and have its impact parameter within $3\sigma$ of the primary
vertex.

SST was studied in a $\Delta m_d$-analysis \cite{SSTPRD}, 
and used in our earlier measurement
of $\sin(2\beta)$. The SST dilution for
the $J/\psi K^0_s$ sample was found
to be $16.6 \pm 2.2$\%  \cite{CDFcp} for events reconstructed in the SVX.
This method has been extended to events out\-side
the SVX coverage (the impact parameter cut is removed);
and we find ${\cal D}^{SST}_{nonSVX} = 17.4 \pm 3.6$\%.

Two ``opposite-side'' taggers, where the
other $b$-hadron signals  
the flavor of the $B^0$, are also used. 
The lepton charge from $b \!\rightarrow\! \ell^-$ decay 
of the other $b$-hadron tags the $B^0$ flavor,
{\it i.e.} $\ell^-$ ($\ell^+$) implies $B^0$ ($\overline{B}{^0}$).
Lepton ($e$ and $\mu$) identification criteria
are applied to all charged tracks\footnote{
Lepton identification limits the tracks to $|\eta| \!<\! 1.0$. Also,
identified conversion electrons are explicitly removed.}
with $p_T$ thresholds of 1.0 (2.0) GeV/c 
for electrons (muons).
The dilution is measured using a $B^+ \rightarrow J/\psi K^+$
sample ($\sim\!1000$ events), 
and we find ${\cal D}^{lep} = 62.5 \pm 14.6$\%.

The other opposite-side method is ``jet-charge.'' 
The tag is a charge average 
of an opposite-side jet. The jet is formed by a mass-clustering
algorithm which starts with ``seed'' tracks of $p_T \!>\! 1.75$ GeV/$c$,
and combines other tracks with  $p_T \!>\! 0.4$ GeV/$c$, up to a cluster
mass approximating the $B$ mass. The $B^0$ decay products are explicitly
excluded from the jet, as are tracks within $\Delta R \!<\! 0.7$ 
of the $B^0$.
If mul\-ti\-ple jet clusters are present, the one most like\-ly to be 
a $b$-jet is chosen based on track impact par\-a\-met\-ers 
and cluster $p_T$.
The jet-charge for a cluster is:
\begin{equation}
       Q_{jet} =
   \frac{{\Sigma}{_{i}}\, q_i
                                p_{ Ti}  (2-T_i)} 
        {{\Sigma}{_{i}}\,
                               {p_{Ti} } (2-T_i)  },
\label{eq:jet-Q}
\end{equation}
where $q_i$ and $p_{Ti}$ are the charge and $p_T$ of the $i$-th
track in the jet with $p_T \!>\! 0.75$ GeV/$c$,
and $T_i$ is the probability that the track
is from the primary vertex.
A $B^0$ ($\overline{B}{^0}$) is implied if $Q_{jet} \!<\! -0.2$ ($>\! 0.2$),
otherwise it is untagged.
The dilution is measured from the $B^+ \!\rightarrow\! J/\psi K^+$
sample to be $23.5 \pm 6.9$\%.

By coincidence, each tagger has an $\epsilon {\cal D}^2$ 
of $\sim\!2$\%. The total  $\epsilon {\cal D}^2$ is $6.3\!\pm\!1.7$\%, 
so our sample of 400 events corresponds to $\sim\!25$
perfectly tagged $J/\psi K^0_s$ decays
plus background.

%%%%%%%%%%%%%%%%%%%%%%%%%%%%%%%%%%%%%%%%%%%%%%%%%%%%%%%%%%%%%%%%%%%%%
\section{\boldmath Extracting $\sin(2\beta)$ \label{sec:sin2beta}}

The three taggers are applied to the sample.
A lepton tends to dominate the jet-charge if a lepton tag is in the jet.
Lepton tagging has low efficiency
but high dilution, so the correlation between lepton and 
jet-charge tags is avoided by dropping the jet-charge tag
if there is a lepton tag. This means each $B^0$ is tagged
at most by two methods. If the tag result for an event by method-$i$ 
is $s_i$ ($s=+1, \; -1, \; 0$ for $B^0$, $\overline{B}{^0}$,
untagged), then the effective dilution for two
tags is ${\cal D}_{ij} = 
|s_i{\cal D}_{i}+ s_j{\cal D}_{i}|/(1+s_i s_j {\cal D}_{i}{\cal D}_{j})$.

An unbinned likelihood fit is performed using
the flavor tags (and the effective ${\cal D}_{ij}$'s),  
$M_N$, and lifetime information from the data;
and it computes the likelihood probability that an event is signal
or background (either prompt or long-lived).
The treatment of the SVX and non-SVX data in the likelihood
is different, but both are part of the same fit.
The $B^0$ lifetime and $\Delta m_d$ values are fixed to 
world averages ($1.54 \pm 0.04$ ps 
and $0.464 \pm 0.018 \,{\rm ps}^{-1}$~\cite{NewPDG}).
The fit also incorporates
allowances for (small) systematic detector biases.

\begin{figure}[t]
{\centering
  \epsfxsize=18.3pc % will enlarge or reduce the postscript figures 
\epsfbox{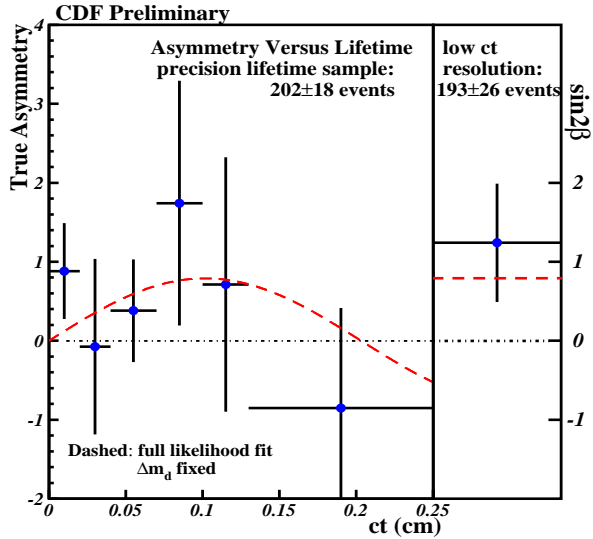}
  \caption{
       The $CP$ asymmetry of the data with the fit
result. The SVX data is shown in proper-time bins on the left, and
a single bin for non-SVX data on the right.
}
\label{fig:sin2beta} }
\end{figure}

The fit yields $\sin(2\beta) \!=\! 0.79^{+0.41}_{-0.44}$
({\it{stat. $+$ syst.}})
for the combined taggers~\cite{CDFNewCP}.
The fit is shown in Fig.~\ref{fig:sin2beta} along with
a dilution weighted average of the sideband-subtracted data.
The result corresponds to 
$0 \!<\! \sin(2\beta)$ for a 93\%
unified frequentist \cite{Feldman} con\-fidence interval.
Although the exclusion of zero has only slightly increased 
from our previous result~\cite{CDFcp},
the uncertainty on $\sin(2\beta)$ is cut in half.

We applied our taggers and fitting
machinery to a sample of $\sim\!450$ $B^0_d \rightarrow J/\psi K^{*0}$ decays
as a cross check.
We find $\Delta m_d \!=\! 0.40 \pm 0.18 \;{\rm ps}^{-1}$, in %accordance 
accor\-dance with the precision of
the $\sin(2\beta)$ analysis.

%%%%%%%%%%%%%%%%%%%%%%%%%%%%%%%%%%%%%%%%%%%%%%%%%%%%%%%%%%%%%%%%%%%%%
\section{Summary and prospects \label{sec:summ}}

We have directly measured $\sin(2\beta)$, and our result provides
evidence for large $CP$ asymmetries in $B^0$ mesons
as expected from indirect determinations,
{\it e.g.} $0.52 \!<\!\sin(2\beta)\!<\! 0.94$ at 95\% CL~\cite{Ali}.

A critical test, however, requires much greater precision.
CDF will attain this in Run II. Starting
in 2000, a 2-year run
should deliver $20\times$ the lum\-inos\-ity ($\sim\!\!2\,$fb$^{-1}$),
and be exploited by a greatly enhanced detector~\cite{CDFup}.
We project $\sim\!10^4$ $J/\psi K^0_s$'s
from dimuons, for a $\sin(2\beta)$ error of about $\pm 0.08$.
Triggering on  $J/\psi \!\rightarrow\! e^+e^-$
may boost the sample by $\sim\!50$\%.
CDF is also working on
a Time-of-Flight system which will aid flavor tagging.
We expect to achieve sensitivities
in the range projected for 
the dedicated $B$ factories.

%------------------------------------------------------------


\begin{thebibliography}{9}

\bibitem{KL_CP} J.H. Christenson {\it et al.}, 
                            \PRL {\bf 13}, 138 (1964). 

\bibitem{CKM} 
% N.~Cabibbo, \PRL {\bf 10},531 (1963);
% Phys. Rev. Lett. {\bf 10}, 531 (1963); 
 M.~Kobayashi and K.~Maskawa,  {\it Prog. Theor. Phys.} {\bf 49}, 652 (1973).

\bibitem{B_CP} J. Bartelt {\it et al.} [CLEO Collab.],
                                         \PRL {\bf 71}, 1680 (1993);
                  F.~Abe {\it et al.} [CDF Collab.]
                                         \PR D {\bf 55} 2546 (1997);
                  K.~Ackerstaff {\it et al.} [OPAL Collab.]
                                      \ZP C {\bf 76}, 401 (1997);
                  G.~Abbiendi {\it et al.}, [OPAL Collab.]
                      CERN-EP/98-195;  %, Submitted to Eur. Phys. J. C
         B.~Petersen [ALEPH Collab.], {\it these proceedings}. %ALEPH

\bibitem{CarterSanda} A.B. Carter and A.I. Sanda,
                                    \PRL {\bf 45}, 952 (1980);
                      I.I. Bigi and A.I. Sanda,
                                    \NP B {\bf 193}, 85 (1981).

\bibitem{CDFcp} F. Abe {\it et al.} [CDF Collab.],
                                   \PRL {\bf 81}, 5513 (1998).

\bibitem{CDFMixSum} A summary of CDF $\Delta m_d$ analyses and the
combined average maybe found in:
    G.~Bauer, {\it Proc. of the 13th
    Topical Conference on Had- ron Collider Physics}, TIFR,
    Mumbai, India, January 1999
    (FERMILAB-Conf-99/042-E). 
    Individual analyses are in:  
      F.~Abe {\it et al.}, % CDF Collaboration,
    \PRL  {\bf 80},   2057 (1998);
    \PR D {\bf 60}, 051101 (1999); %dimu
    FERMILAB-Pub-99/019-E; %jet-Q
    FERMILAB-Pub-99/210-E. %fumi lep-D*   hep-ex/9907053 T. Affolder !!

\bibitem{Artuso} M. Artuso, {\it these proceedings}.

\bibitem{Gronau} M. Gronau, A. Nippe, and J. Rosner,
                    \PR D {\bf 47}, 1988 (1993); 
                 M.~Gronau and J.~Rosner, {\it ibid.} {\bf 49}, 254 (1994).
 
\bibitem{Dejan} G. Bauer [CDF Collab.], {\it these proceedings}
       (FERMILAB-Conf-99/227-E);
       D.~Vucinic, Ph.D. dissertation, 
       Massachusetts Institute of Technology, 1999.

\bibitem{SSTPRD} F. Abe {\it et al.} [CDF Collab.],
                                   \PR D {\bf 59}, 032001 (1999).\\

\bibitem{NewPDG} Particle Data Group, C. Caso {\it et al.},
                      {\it Eur. Phys. J.} C {\bf 3}, 1 (1998).

\bibitem{CDFNewCP} T. Affolder {\it et al.} [CDF Collab.], 
 submitted to    \PR D 
                (FERMILAB-Pub-99/225-E).

\bibitem{Feldman} G.J. Feldman and R.D. Cousins,
                                   \PR D {\bf 57}, 3873 (1998).

\bibitem{Ali} A. Ali and D. London, {\it Eur. Phys. J.} C {\bf 9}, 687 (1999).

\bibitem{CDFup} CDFII Collaboration,  
                FERMILAB-Pub-96/ 390-E. 

\end{thebibliography}
\end{document}